\documentclass[10pt,aps,prl,twocolumn,superscriptaddress,nobibnotes,nodoi,noeprint]{revtex4-2}

\usepackage{float}
\usepackage{amsmath,physics,bm,float}
\usepackage{soul}
\usepackage[colorlinks=true,citecolor=blue,urlcolor=blue,linkcolor=blue]{hyperref}
\usepackage{amssymb}
\usepackage{natbib}
\usepackage{upgreek}
\usepackage{mathtools}
\usepackage{siunitx}
\usepackage[normalem]{ulem}
\usepackage{graphicx}
\usepackage{ragged2e}
\usepackage{amsmath}
\usepackage{placeins}



\usepackage[mathlines]{lineno}
\let\oldequation\equation\let\oldendequation\endequation
\renewenvironment{equation}{\linenomathNonumbers\oldequation}{\oldendequation\endlinenomath}
\let\oldalign\align\let\oldendalign\endalign
\renewenvironment{align}{\linenomathNonumbers\oldalign}{\oldendalign\endlinenomath}

\begin{document}
	
   \title{Flocking as a second-order phase transition in self-aligning active crystals}

    \author{Marco Musacchio}
	\affiliation{
		Institut f{\"u}r Theoretische Physik II: Weiche Materie,
		Heinrich-Heine-Universit{\"a}t D{\"u}sseldorf, Universit{\"a}tsstra{\ss}e 1,
		D-40225 D{\"u}sseldorf, 
		Germany}
    
	\author{Alexander P.\ Antonov}
	\affiliation{
		Institut f{\"u}r Theoretische Physik II: Weiche Materie,
		Heinrich-Heine-Universit{\"a}t D{\"u}sseldorf, Universit{\"a}tsstra{\ss}e 1,
		D-40225 D{\"u}sseldorf, 
		Germany}

	
	
	\author{Hartmut L{\"o}wen}
	\affiliation{
		Institut f{\"u}r Theoretische Physik II: Weiche Materie,
		Heinrich-Heine-Universit{\"a}t D{\"u}sseldorf, Universit{\"a}tsstra{\ss}e 1,
		D-40225 D{\"u}sseldorf, 
		Germany}

    \author{Lorenzo Caprini}
	\email{lorenzo.caprini@uniroma1.it}
	\affiliation{
		Physics Department, University of Rome La Sapienza, P.le Aldo Moro 5, IT-00185 Rome, Italy}
	
	\date{\today}
        
	\begin{abstract} 
       \noindent 
       We study a two-dimensional crystal composed of active units governed by self-alignment. This mechanism induces a torque that aligns a particle’s orientation with its velocity and leads to a phase transition from a disordered to a flocking crystal. Here, we provide the first microscopic theory that analytically maps the crystal dynamics onto a Landau-Ginzburg model, in which the velocity-dependent effective free energy undergoes a transition from a single-well shape to a Mexican-hat profile. As confirmed by simulations, our theory quantitatively predicts the transition point and characteristic spatial velocity correlations. The continuous change of the order parameter and the diverging behavior of the analytically predicted correlation length imply that flocking in self-aligning active crystals is a second-order phase transition. These findings provide a theoretical foundation for the flocking phenomenon observed experimentally in active granular particles and migrating cells.  
       \end{abstract}
	
	\maketitle


Living systems across all length scales~\cite{marchetti2013hydrodynamics, Elgeti2015} display self-organization and collective phenomena, ranging from the flocking of animals to the swarming of bacteria~\cite{kearns2010field} and the collective migration of cells~\cite{alert2020physical}. Their natural behaviors have inspired the design of synthetic active matter units~\cite{bechinger2016active}, such as Quincke rollers~\cite{bricard2013emergence, geyer2019freezing} or active robots~\cite{turgut2008self, vasarhelyi2018optimized, Agrawal2020}. These systems often exhibit swarming or flocking phenomena~\cite{deseigne2010collective, kumar2014flocking, viragh2014flocking, chen2023molecular, das2024flocking} and have the potential to inspire innovative technologies~\cite{brambilla2013swarm}.

Over the last two decades, these perspectives have posed theoretical challenges in identifying the minimal ingredients required to reproduce these collective behaviors and in developing predictive theories.
The observed collective movement is often modeled through explicit alignment interactions that couple the velocities of different particles~\cite{vicsek2012collective, chate2008modeling, ihle2011kinetic, solon2013revisiting, barberis2016large, shankar2017topological, levis2019activity, casiulis2020velocity, chatterjee2021inertia, geiss2022information, kreienkamp2022clustering, giraldo2025active}. This approach is inspired by the Ising and XY models~\cite{binney1992theory}, originally introduced to describe the phase transition from paramagnetic to ferromagnetic materials. 
These velocity-aligning interactions have successfully described flocking in animals such as birds~\cite{cavagna2014bird} and insects~\cite{cavagna2017dynamic}, interpreting
this phenomenon as a phase transition.

In contrast to animals, active systems like migrating cells and swarming robots are described by alternative aligning mechanisms. These systems are typically polar and are often characterized by a coupling between translational and rotational motion, commonly referred to as self-alignment~\cite{baconnier2024self}. This mechanism acts at the level of individual particles as an internal torque that tends to align the particle's orientation with its velocity.
After being proposed to describe cells~\cite{szabo2006phase} and epithelial tissues~\cite{malinverno2017endocytic, barton2017active, giavazzi2018flocking, shen2025flocking}, self-alignment has been identified as the key mechanism governing the dynamics of polar active granular particles, self-propelling due to an internal motor~\cite{giomi2013swarming, deblais2018boundaries, dauchot2019dynamics, leoni2020surfing, tapia2021trapped, siebers2023exploiting, altshuler2024environmental} or a vibrating plate~\cite{aranson2007swirling, kudrolli2008, Koumakis2016, scholz2018rotating, lopez2022chirality, antonov2024inertial, antonov2025self}.
In these cases, self-alignment arises from a misalignment between the geometric center and the center of mass of the particle, generating a torque that aligns the particle's orientation with its velocity. 

Self-alignment has been identified as the key mechanism inducing the flocking transition in polar granular systems~\cite{lam2015self, casiulis2024geometric}, as experimentally and numerically explored. 
At higher densities, such as self-aligning active liquids~\cite{shimoyama1996collective, henkes2011active, canavello2024polar, musacchio2025self}, glasses~\cite{paoluzzi2024flocking} and crystals~\cite{ferrante2013elasticity, ferrante2013collective, baconnier2022selective, kinoshita2025collective}, the system undergoes a transition from a homogeneous disordered state to a homogeneous flocking one. These flocking phases emerge when the strength of the self-aligning torque dominates over the disruptive effects of translational and rotational noise.

However, despite systematic numerical investigations and its demonstrated relevance in experiments, a microscopic theory that fully captures the spontaneous collective motion induced by self-alignment is still lacking. Consequently, fundamental questions -- such as whether the self-alignment-induced flocking transition is of first or second order -- remain open.

\begin{figure}[t] 
    \centering
    \includegraphics[width=0.85\columnwidth]{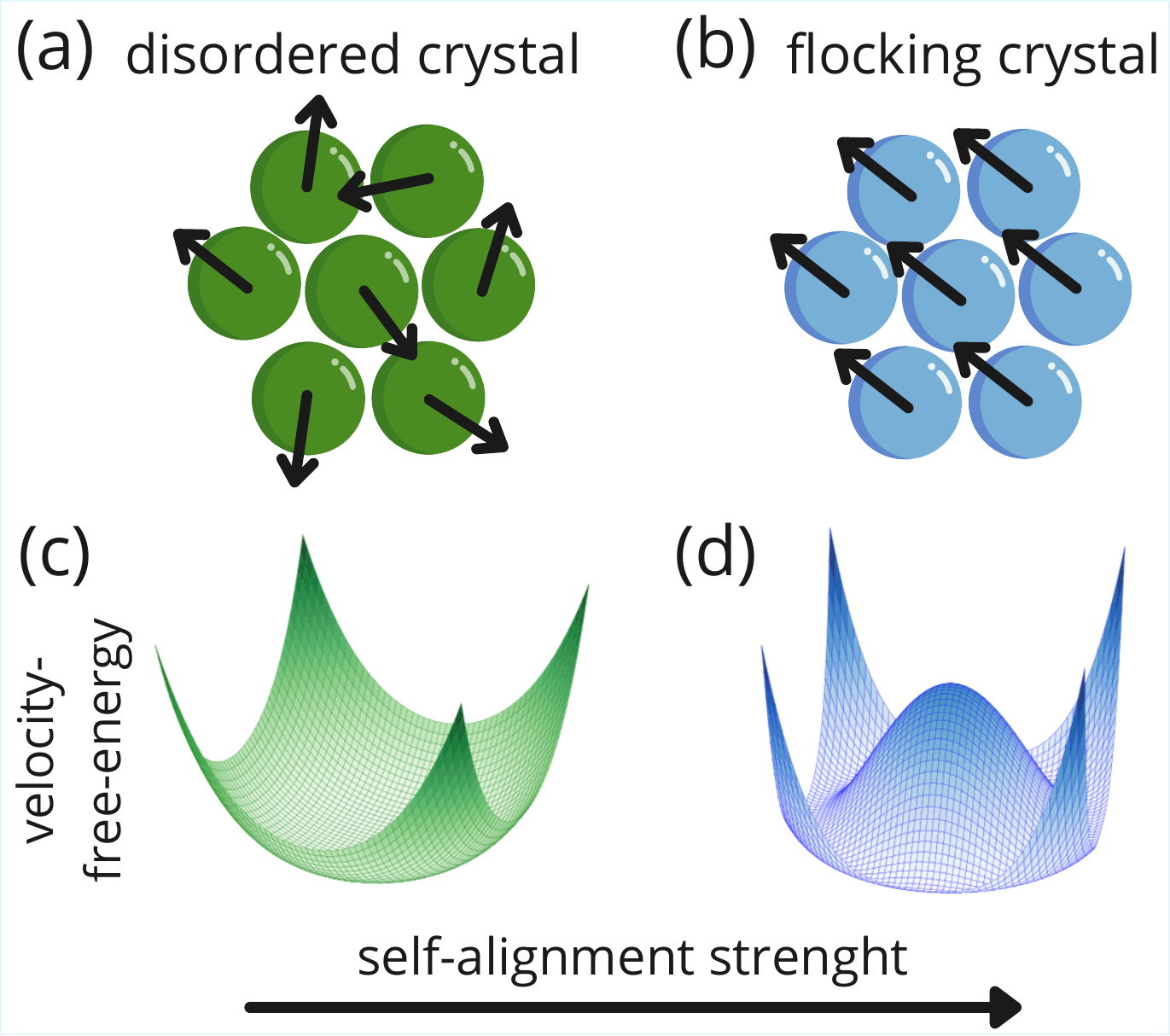}
    \caption{\textbf{Phase transition from a disordered to a flocking state in a self-aligning active crystal.} (a)–(b) Illustrations showing the disordered phase (a) and the flocking phase (b), where particle velocities are indicated by black arrows. (c)–(d) Corresponding Landau-Ginzburg velocity-free-energy profiles, transitioning from a single-well to a Mexican-hat-like profile.}
    \label{fig:figura1}
\end{figure}

We fill this gap by introducing, to the best of our knowledge, the first microscopic theory describing the flocking transition in dense self-aligning active matter. We map the dynamics of a self-aligning active crystal onto a Landau-Ginzburg model with a velocity-dependent free energy, which undergoes a second-order phase transition from a single-well potential (Fig.~\ref{fig:figura1}(a)) (corresponding to the disordered phase (Fig.~\ref{fig:figura1}(c))) to a Mexican-hat profile (Fig.~\ref{fig:figura1}(d)) (corresponding to the flocking phase (Fig.~\ref{fig:figura1}(b))). The theory yields an exact prediction for the transition point as a function of self-alignment strength and persistence length and accurately captures both the spatial structure of velocity correlations and the associated correlation length.


We consider a two-dimensional system of $N$ interacting active Brownian particles subject to a self-alignment mechanism. ``Active'' means that each particle moves persistently with a velocity aligned with its orientation, while ``self-alignment'' implies that this orientation tends to align with the particle’s velocity due to a torque. Motivated by granular experiments, the particles can be described by inertial translational dynamics and overdamped rotational dynamics. The equations of motion for a particle with position $\mathbf{x}_i$, velocity $\mathbf{v}_i=\dot{\mathbf{x}}_i$, and orientational angle $\theta$ are given by
\begin{subequations}
    \begin{align}
    m\dot{\mathbf{v}}_i &= -\gamma \mathbf{v}_i + \gamma v_0\mathbf{\hat{n}}_i + \mathbf{F}_i + \gamma\sqrt{2 D_t}\,\boldsymbol{\xi}_i \,, \label{eq:eulero_pos} \\
     \gamma_r \dot{\theta}_i  &=  \mathbf{T}^{sa}_i \cdot \hat{\mathbf{e}}_z + \gamma_r \sqrt{2 D_r} \eta_i  \,.
     \label{eq:eulero_ang}
\end{align}
\end{subequations}
The terms $\gamma$ and $\gamma_r$ denote the translational and rotational friction coefficients, $D_t$ and $D_r$ are the corresponding diffusion coefficients, while $\boldsymbol{\xi}_i$ and $\eta_i$ represent independent Gaussian white noises with zero mean and unit variance. Each particle, of mass $m$ is propelled at constant speed $v_0$ along the direction $\hat{\mathbf{n}}_i = (\cos\theta_i, \sin\theta_i)$, described by the orientational angle $\theta_i$. This is subject to a deterministic torque $\mathbf{T}^{\text{sa}}_i=\beta\, (\hat{\mathbf{n}}_i \times \mathbf{v}_i)$ which aligns $\mathbf{n}_i$ to $\mathbf{v}_i$. The strength of the self-alignment is controlled by the parameter $\beta$, which sets the typical distance $\gamma_r/\beta$ traveled by an active particle before its orientation aligns with its velocity. 
Finally, the force $\mathbf{F}_i$ models the pure repulsive interactions between the particles and is derived from a repulsive Weeks-Chandler-Andersen potential: $\mathbf{F}_i = -\nabla_i U_{\rm tot}$, with $U_{\rm tot} = \sum_{i < j} U(\lvert \mathbf{r}_i - \mathbf{r}_j \rvert)$ and $U(r) = 4\epsilon[(\sigma/r)^{12} - (\sigma/r)^6]$. Here, $\epsilon$ sets the energy scale and $\sigma$ represents the particle diameter. In our simulations, we set $D_t = 0$, since in active matter this term is typically lower than the effective diffusion induced by activity.

Self-alignment introduces an additional typical time, $\gamma_r/(\beta v_0)$, i.e.\ the time needed by the orientation to align with the velocity. This time competes with the other typical times governing the dynamics, i.e.\ the particle persistence time, $\tau = 1/D_r$, and the translational inertial time, $\tau_d = m/\gamma$.
Simulations are performed by rescaling lengths by the particle diameter $\sigma$ and time by $\tau$. In this way, the dynamics are governed by several dimensionless parameters: the P\'eclet number $\text{Pe} = v_0/(D_r \sigma)$, which quantifies the persistence length of the particle compared to its size; the reduced mass $M = D_r m/\gamma$ setting the relevance of inertia and chosen low as $M=10^{-3}$;  the reduced self-alignment strength $B = \beta \sigma / \gamma_r$ which compares the self-aligning length with the particle size; and, finally, the reduced interaction strength, given by $\sqrt{\epsilon/m}/(D_r \sigma)$.

\begin{figure}[t] 
    \centering
    \includegraphics[width=\columnwidth]{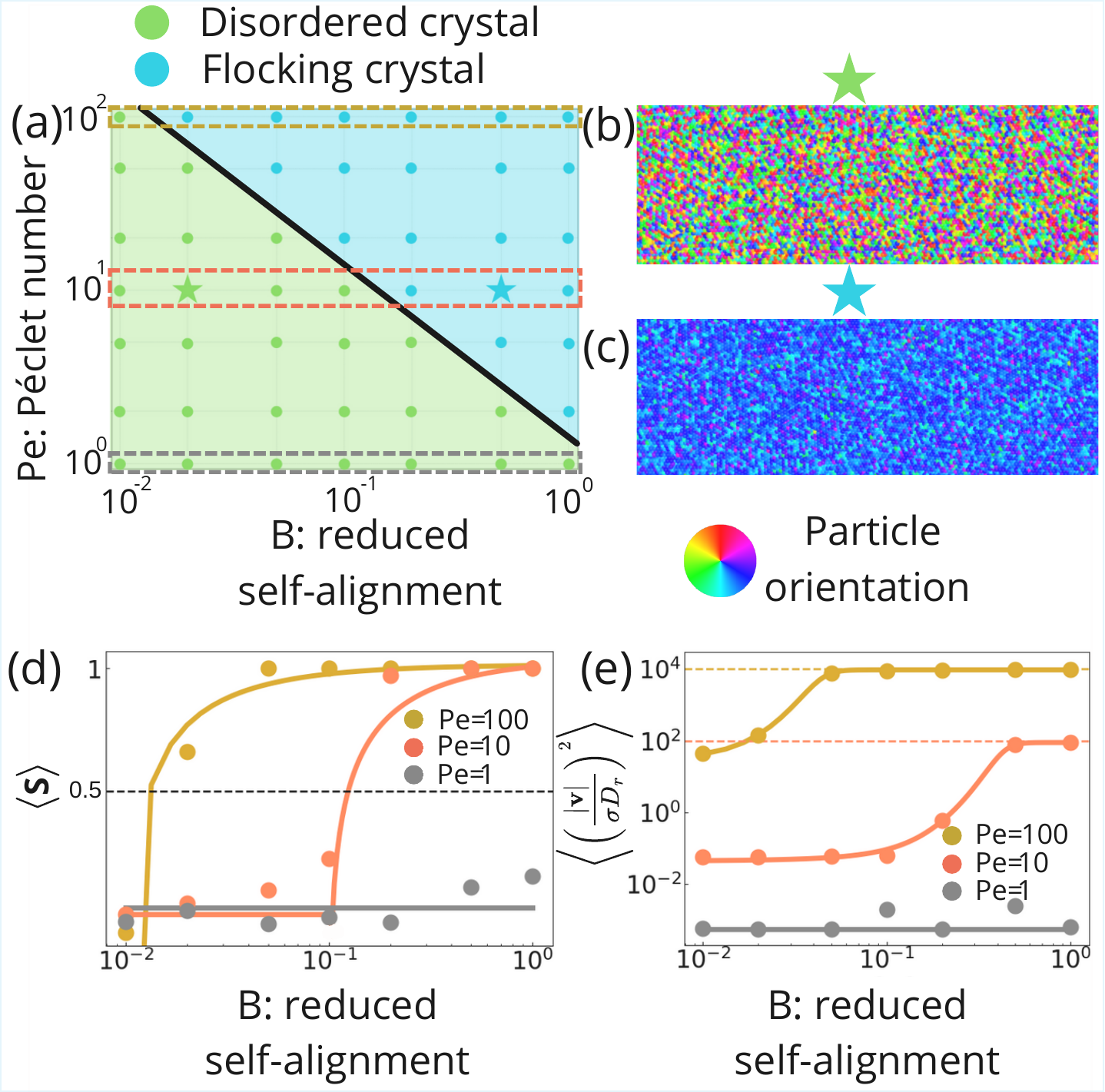}
    \caption{\textbf{Disorderd-flocking phase diagram.} (a) Phase diagram in the plane of the P\`eclet number $\text{Pe} = v_0/(D_r \sigma)$ and the reduced self-alignment strength $B = \beta \sigma / \gamma_r$, showing disordered (green) and flocking (blue) crystals. (b)–(c) Simulation snapshots corresponding to the two stars in the phase diagram. Particles' colors indicate the orientation of the particle velocity. The black line marks the transition for $B\to B_c=\text{Pe}^{-1}$ ($\beta_c/\gamma_r= 1/(v_0 \tau)$) corresponding to the theoretical prediction~\eqref{eq:transitionpoint}. 
    (d) Average polarization $\langle \mathbf{S}\rangle$ and (e) average squared velocity $\langle v^2 \rangle$ as functions of $B$ for different P\`eclet numbers, in correspondence of the dashed rectangles in (a). The dashed black line In (d) marks 0.5, chosen as the threshold from a disordered ($\langle \mathbf{S}\rangle\sim 0$) to a flocking state ($\langle \mathbf{S}\rangle\sim 1$). Colored lines in (d) are obtained by plotting the theoretical prediction~\eqref{eq:S_prediction_main} with an additive constant accounting for finite size effects visible in the disordered phase. Colored, solid lines in (e) are guides for the eyes, while colored, dashed horizontal lines in (e) are drawn in correspondence with the free-particle speed - This is not reached for $\text{Pe} = 1$. The remaining dimensionless parameters of the simulations are $M = 10^{-4}$, $\sqrt{\epsilon/m}/(D_r \sigma)=10^2$, and $\Phi = N \pi \sigma^2 / 4L^2 = 1.1$.
  }
    \label{fig:figura2}
\end{figure}

We focus on an active crystal composed of self-aligning active Brownian particles. To guarantee a homogeneous crystal configuration, simulations of Eq.~\eqref{eq:eulero_ang} are performed in a two-dimensional square box, compatible with a hexagonal crystal, of size $L$ with periodic boundary conditions and a high packing fraction value $\Phi=N\sigma^2 \pi/(4 L^2)=1.1$.
As known in previous studies~\cite{paoluzzi2024flocking}, when the self-alignment strength $\beta$ is sufficiently large, compared to the particle persistence length $v_0/D_r$ (Fig.~\ref{fig:figura2}~(a)), the system shows a transition from a disordered (Fig.~\ref{fig:figura2}~(b)) to a flocking phase (Fig.~\ref{fig:figura2}~(c)). The flocking transition is identified by monitoring the velocity polarization $\langle S \rangle = 1/N\left\langle \left| \sum_i \mathbf{v}_i/|\mathbf{v}_i| \right| \right\rangle$ (Fig.~\ref{fig:figura2}(d)), and the average squared velocity $\langle v^2 \rangle$ (Fig.~\ref{fig:figura2}(e)), as function of the reduced self-alignment strength $B$. In a flocking configuration, $\langle S \rangle$ is close to the unit while $\langle v^2 \rangle$ approaches the single-particle velocity $v_0$. By contrast, in the disordered phase, $\langle S \rangle \ll 1$ while $\langle v^2 \rangle \ll v_0$. The flocking transition line is identified when $\langle S \rangle>0.5$.

The main result of this paper is the derivation of a microscopic theory that predicts flocking in self-aligning active crystals and interprets this phenomenon as a second-order phase transition.

By applying the time-derivative to the dynamics~\eqref{eq:eulero_pos}, and eliminating $\dot{\mathbf{n}}_i$ using Eq.~\eqref{eq:eulero_ang}, we can obtain an effective evolution equation for the particle velocity (see Supplemental Material - SM) which is exact for $m/(\gamma\tau)\to 0$, i.e.\ the regime numerically evaluated. This theoretical mapping has already been applied in active systems without self-alignment~\cite{marconi2015towards, fodor2016far}, for instance to predict the wall accumulation typical of active Brownian particles~\cite{maggi2015multidimensional, das2018confined}.
In a crystal, we adopt the lattice approximation by assuming that particles occupy fixed positions on a lattice, such that the net force on each particle is balanced, $\mathbf{F}_i = 0$. Within this approximation, the dynamics simplify (see SM for details on the derivation) and are given by
\begin{equation}
    \begin{aligned}
    \dot{\mathbf{v}}_i  \approx &- \sum_{j=1}^{n_i} \frac{\mathcal{K}}{\gamma} \left(\mathbf{v}_i -  \mathbf{v}_j \right) + v_0\sqrt{\frac{2}{\tau}} \boldsymbol{w}_i\\
    &- \mathbf{v}_i\left(\frac{1}{\tau}-\frac{\beta}{\gamma_r} v_0 \right)
    - \frac{\beta}{\gamma_r v_0^2} \mathbf{v}_i |\mathbf{v}_i|^2 \,,
    \end{aligned}
    \label{eq:lattice_approx}
\end{equation}
where the sum $\sum_j^{n_i}$ is restricted to the six neighbors of the particle $i$ and $w_i$ is a white noise vector with zero average and unit variance. The constant term $\mathcal{K}$ is related to the Hessian matrix of the interaction potential, i.e.\ its second derivative, and consequently, $\mathcal{K}$ mainly depends on the lattice structure (see SM for details). 
In Eq.\eqref{eq:lattice_approx}, the first line includes a noise contribution and an effective velocity-dependent interaction term $\mathcal{K} \left(\mathbf{v}_i - \mathbf{v}_j \right)$, which acts as a discrete Laplacian and promotes alignment of the velocities among neighboring particles.
Compared to non-aligning active particles\cite{marconi2016velocity, caprini2020spontaneous}, each $\mathbf{v}_i$ is additionally subject to a single-body, velocity-dependent force, as shown in the second line of Eq.~\eqref{eq:lattice_approx}.

To interpret the results, we switch from a particle-based description to a continuum approach $\mathbf{v}_i \to \mathbf{v}(\mathbf{r})$ for the velocity field.
In this way, $\mathbf{v}(\mathbf{r})$ evolves as
\begin{equation}
\dot{\mathbf{v}}(\mathbf{r})=-\frac{1}{\tau}\frac{\delta}{\delta \mathbf{v}(\mathbf{r})} \mathcal{F}[\mathbf{v}(\mathbf{r})] +  v_0\sqrt{\frac{2}{\tau}} \boldsymbol{w}(\mathbf{r})
\end{equation} 
where $\boldsymbol{w}(\mathbf{r},t)$ is a white noise with zero average for every position $\mathbf{r}$, such that $\langle \boldsymbol{w}(\mathbf{r}, t)\boldsymbol{w}(\mathbf{r'}, t')\rangle=\delta(t-t')\delta(r-r')$. The term $\mathcal{F}=\mathcal{F}[\mathbf{v}(\mathbf{r})]$ is an effective free-energy functional $\mathcal{F}_{LG}=\mathcal{F}_{LG}[\mathbf{v}(\mathbf{r})]$ which uniquely depends on the velocity field:
\begin{equation}
\label{eq:free_energy}
\mathcal{F}[\mathbf{v}] =\frac{3\tau\sigma^2\mathcal{K}}{2 \gamma}\left(\nabla \mathbf{v}\right)^2 + \left(1 - \frac{v_0\tau\beta}{\gamma_r}\right)\frac{|\mathbf{v}|^2}{2} + \frac{\tau\beta}{\gamma_r} \frac{|\mathbf{v}|^4}{4 v_0^2}\,.    
\end{equation}
This free energy has a Landau-Ginzburg shape and is characterized by the usual mass term $\propto \mathbf{v}^2$, interaction $\propto \mathbf{v}^4$, and gradient term $\propto (\nabla\mathbf{v})^2$ with coefficients depending on the persistence length $v_0\tau$ and self-alignment strength $\beta$.
The gradient term favors the homogeneous phase penalizing spatial changes of $\mathbf{v}(\mathbf{r})$ and is generated by the discrete Laplacian contribution proportional to $\propto \mathcal{K}$ in Eq.~\eqref{eq:lattice_approx}.
  
By contrast, the mass and interaction terms are obtained by integrating over $\mathbf{v}$ the linear and cubic contributions in Eq.~\eqref{eq:lattice_approx}. 
Specifically, the mass term changes sign when $\beta$ exceeds the critical value $\beta_c$, given by
\begin{equation}
\label{eq:transitionpoint}
    \frac{\beta_c}{\gamma_r} = \frac{1}{v_0 \tau} \,,
\end{equation}
corresponding to $B_c=\text{Pe}^{-1}$. As a result, when the typical self-alignment length is larger than the persistence length, $\beta<\beta_c$ ($B<B_c$), the free energy is peaked at zero implying the stability of the disordered phase. By contrast, in the opposite regime $\beta>\beta_c$ ($B>B_c$), the free energy has a mexican-hat shape consistently with a flocking phase where rotational symmetry is broken. In this regime, the system displays a flocking behavior, characterized by a non-vanishing velocity field.
The predicted transition point $\beta=\beta_c$ (or $B=B_c$) is in agreement with our numerical findings (black line in Fig.~\ref{fig:figura2}~(a)). 
To the best of our knowledge, this is the first theoretical microscopic prediction explaining the flocking behavior in self-aligning active systems.

From the expression~\eqref{eq:free_energy}, we can predict the polarization $\langle \mathbf{S} \rangle$ as a function of the self-alignment strength $\beta$ (see SM). In a disordered phase $\langle \mathbf{S} \rangle\approx 0$ or reaches small values $\ll 1$ because of finite size-effects (Fig.~\ref{fig:figura2}~(d)). By contrast, in the flocking state, $\langle \mathbf{S} \rangle$ can be estimated from the modulus of the velocity value which minimizes the effective free-energy~\eqref{eq:free_energy}
\begin{equation}
\label{eq:S_prediction_main}
    |\mathbf{S}| 
    \approx \left(  \frac{\gamma_r}{\beta}\right)^{1/2}\left(\frac{\beta}{\gamma_r} - \frac{1}{v_0\tau} \right)^{1/2} \propto \sqrt{\frac{{B-B_c}}{B}} \,.
\end{equation}
As confirmed numerically, $\langle \mathbf{S} \rangle$ increases continuously with the square root of the reduced self-alignment strength $B$ (Fig.~\ref{fig:figura2}~(d)), displaying a behavior reminiscent of the magnetization curve in the Ising model~\cite{binney1992theory}. This represents evidence that flocking in a self-aligning crystal undergoes a second-order phase transition.

\begin{figure}[t] 
    \centering 
    \includegraphics[width=\columnwidth]
    {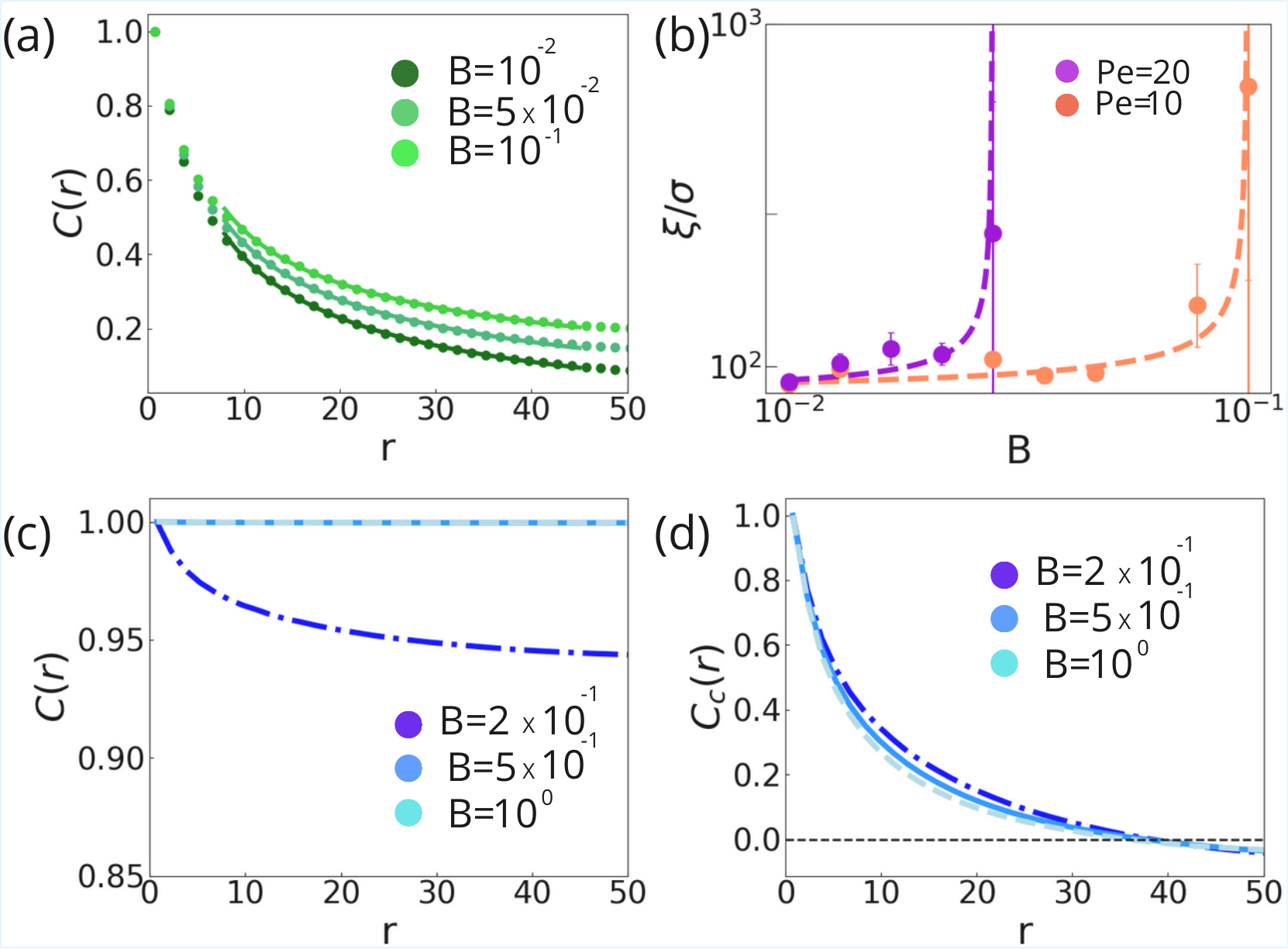}
    \caption{\textbf{Spatial velocity correlations and correlation length.} (a) Spatial velocity correlations for different values of the reduced self-alignment strength $B$ before the flocking transition at $\text{Pe} = 10$. The dashed lines represent simulation data, while the solid curves correspond to the theoretical prediction, obtained by fitting the function $f(r) = a\, e^{-r/\xi}/\sqrt{r}$, where $a$ and $\xi$ are the fit parameters. (b) correlation length $\xi$, obtained from the previous fits, is plotted as a function of $B$ for $\text{Pe} = 20$ and $\text{Pe} = 10$. The correlation length $\xi$ exhibits a diverging behavior as the system approaches the critical point $\beta \to \beta_c = \gamma_r / (v_0 \tau)$, corresponding to $B \to B_c = \text{Pe}^{-1}$, in agreement with theoretical predictions. The critical points are marked with vertical dashed lines, shown in purple for $\text{Pe} = 20$ and in orange for $\text{Pe} = 10$.
 (Eq.~\eqref{eq:lambda_theory}). (c)–(d): Spatial velocity correlations $C(r)$ and connected spatial velocity correlations $C_c(r)$, after the flocking transition for different values of $B$ at $\text{Pe} = 10$. The remaining dimensionless parameters of the simulations are $M = 10^{-4}$, $\sqrt{\epsilon/m}/(D_r \sigma)=10^2$, and $\Phi = N \pi \sigma^2 / 4L^2 = 1.1$.}
    \label{fig:figura3}
\end{figure}

Consistently with a Landau-Ginzburg model, the spatial velocity correlations 
$C(r) = \langle \sum_{ij} \mathbf{v}_i \cdot\mathbf{v}_j \delta(r-r_{ij})/\sum_{ij} \delta(r-r_{ij})\rangle$ - with $r_{ij}=|\mathbf{r}_i -\mathbf{r}_{j}|$ -
have a spatial exponential shape $\sim e^{-r/\lambda}/\sqrt{r}$ before the transition point (Fig.~\ref{fig:figura3}~(a)). 
Here, $\lambda$ corresponds to the correlation length of $C(r)$ which can be analytically predicted as
\begin{equation}
    \label{eq:lambda_theory}
    \lambda = 
    \frac{\sigma\sqrt{3\tau\mathcal{K}}}{\sqrt{\gamma}\left(1-\frac{v_0 \tau \beta}{\gamma_r}\right)^{1/2}}
    = \sigma\sqrt{\frac{3\tau\mathcal{K}}{\gamma}}\frac{1}{\left(1-B \text{Pe}\right)^{1/2}}\,.
\end{equation}
The correlation length $\lambda$ diverges at the transition point, occurring when $\beta/\gamma_r \to \beta_c = 1/(v_0 \tau)$, i.e., when $B \to B_c = \text{Pe}^{-1}$. This divergence is evident as $\lambda$ exceeds the system size (Fig.~\ref{fig:figura3}(b)), providing further confirmation that flocking in dense self-aligning systems is a second-order phase transition.
After the transition, the correlation $C(r)$ shows a trivial flat profile because each particle velocity is determined by the average value $\langle \mathbf{v} \rangle$ (Fig.~\ref{fig:figura3}~(c)). By following Cavagna and Giardina~\cite{cavagna2018physics}, we monitor the connected correlation function $C_c(r) = \langle \sum_{ij} \delta\mathbf{v}_i \cdot \delta\mathbf{v}_j \delta(r-r_{ij})/\sum_{ij} \delta(r-r_{ij})\rangle$ - with $r_{ij}=|\mathbf{r}_i -\mathbf{r}_{j}|$, obtained by considering the deviation of the velocity field from its spatial average $\delta \mathbf{v}_i=\mathbf{v}_i - \langle\mathbf{v}\rangle$.
This function $C_c(r)$ crosses zero and reaches negative values (Fig.~\ref{fig:figura3}~(d)) as in previous studies on the flocking behaviors of birds~\cite{cavagna2018physics}. Consistently, our system shows the typical scale-free properties of a flocking state: the curves for $C_c(r)$ collapse for different values of the parameters being uniquely determined by the system size.

Our study provides the first microscopic theory for the flocking phenomenon in self-aligning active crystals by establishing a theoretical link between self-aligning active matter and the well-established Ginzburg–Landau model used to describe order-disorder phase transitions. The self-alignment strength induces a sign change in the mass term of the effective velocity-dependent free energy, giving rise to a symmetry breaking in the rotational symmetry of the system and, consequently, Goldstone modes.  
In contrast to Vicsek-like models~\cite{baglietto2009nature, aldana2007phase, chate2007comment}, our velocity Ginzburg–Landau theory predicts that flocking in self-aligning active crystals is a second-order phase transition. This is evidenced by the continuous change of the order parameter - the velocity polarization - and by the divergence of the correlation length of the spatial velocity correlations.

Compared to Vicsek-like models, flocking is observed here without an explicit coupling between different particle velocities. However, to achieve this global movement, both repulsive interactions and self-alignment are crucial, as evidenced by our theory: in the absence of self-alignment, the mass term never changes sign, and the disordered phase remains always stable. Indeed, collective motion in crystals composed of non-aligning active Brownian particles cannot be observed~\cite{bialke2012crystallization, digregorio2018full, henkes2020dense, caprini2023entropons, keta2024emerging}, except in the limit of infinite persistence length~\cite{caprini2023flocking}. Consequently, our microscopic theory justifies the formulation of a hydrodynamic description in self-aligning active matter and could inspire a more systematic theoretical study, for instance through a Toner-Tu theory~\cite{toner2005hydrodynamics}. This work thus offers a solid theoretical foundation for understanding collective behavior in high-density active granular matter exhibiting swarming or flocking, as well as in migrating biological systems, such as cell monolayers or living tissues, which are often modeled through self-alignment mechanisms.

\paragraph*{Acknowledgments:}
HL acknowledges support by the Deutsche Forschungsgemeinschaft (DFG) through the SPP 2265, under grant numbers LO 418/25. 
LC acknowledges financial support from the University of Rome Sapienza, under the project Ateneo 2024 (RM124190C54BE48D).

    \bibliographystyle{apsrev4-1}

    \bibliography{bib.bib}





\end{document}